# Energy Efficiency of the IEEE 802.15.4 Standard in Dense Wireless Microsensor Networks: Modeling and Improvement Perspectives

Bruno Bougard[13], Francky Catthoor[13] Denis C. Daly[2], Anantha Chandrakasan[2] Wim Dehaene[3]

[1]*IMEC, Leuven Belgium*

[2]*Department of EECS, Massachusetts Institute of Technology, Cambridge, MA 02139*

[3]*Department of EE, K.U. Leuven, Belgium*

*{bougardb, catthoor}@imec.be* *{ddaly, anantha}@mit.edu* *wim.dehaene@esat.kuleuven.ac.be*

## Abstract

**Wireless microsensor networks, which have been the topic of intensive research in recent years, are now emerging in industrial applications. An important milestone in this transition has been the release of the IEEE 802.15.4 standard that specifies interoperable wireless physical and medium access control layers targeted to sensor node radios. In this paper, we evaluate the potential of an 802.15.4 radio for use in an ultra low power sensor node operating in a dense network. Starting from measurements carried out on the off-the-shelf radio, effective radio activation and link adaptation policies are derived. It is shown that, in a typical sensor network scenario, the average power per node can be reduced down to 211µW. Next, the energy consumption breakdown between the different phases of a packet transmission is presented, indicating which part of the transceiver architecture can most effectively be optimized in order to further reduce the radio power, enabling self-powered wireless microsensor networks.**

## 1. Introduction

Wireless microsensor networks are autonomous networks for monitoring purposes, ranging from short-range, potentially *in vivo* health monitoring [1] to wide-range environmental surveillance [2]. Thanks to the tremendous range of applications they will enable, wireless microsensor networks have received a great deal of attention in recent years. Designing such a network and more specifically the protocols to support its functioning, is a challenging task. Despite the wide variety of applications, all sensors networks face similar constraints [3]:

***Density*** High-end microsensor networks are expected to have a density of approximately 20 nodes/m$^3$. Hence, the medium access control layer (MAC) should be able to accommodate several hundred to thousand nodes.

***Distributed traffic*** Due to their high node density, wireless sensor networks must have a high capacity. However, the data rate requirements per node are low (<10 kbps). This results in a very low radio duty cycle.

***Energy*** Microsensor nodes are required to be small and autonomous. Their small form factor limits the amount of energy that can be stored in batteries. Furthermore, the density of the network as well as the environment where nodes are deployed often prohibits periodic replacement of the batteries. An existing goal is for a microsensor node to have an average power on the order of 100µW, which would allow the device to obtain its power from the environment by energy scavenging [4].

Wireless microsensor network research in recent years has strived to design radio circuitry and transmission protocols to meet these novel constraints [5,6] and it is expected that results from this research will soon emerge in industrial applications. An important milestone in this transition has been the release of the IEEE 802.15.4 standard [7] that specifies interoperable physical and medium access control layers targeted to sensor node radios.

In [8], the performance of the 802.15.4 standard in terms of throughput and energy efficiency is assessed based on simulation. However, this work focuses on a scenario with few nodes and low load, which diverges significantly from the conditions encountered in wireless microsensor networks. In this paper, we evaluate the potential of the standard for use in an ultra low power sensor node operating in the aforementioned dense network conditions.

Section 2 provides a short description of the 802.15.4 standard and outlines a commercial radio implementing the standard, the Chipcon CC2420 [9]. Measurement results carried out on the off-the-shelf component are presented in section 3. Next, in section 4, we develop an energy-aware activation policy for the radio and model the resulting average power consumption, as well as the corresponding transmission reliability. In section 5, it is shown how the model can be used to optimize the energy efficiency in a dense microsensor network scenario. Finally, the energy consumption breakdowns between the different phases of a packet transmission and the different states of the radio are presented, indicating which part of the transceiver architecture can most effectively be optimized in order to further reduce the radio power consumption and thus enable self-sustained wireless microsensor networks.

## 2. The IEEE 802.15.4 standard and its implementation

### Physical layer

An IEEE 802.15.4 compliant radio can operate in 16 channels in the 2450MHz ISM band, 10 channels in the 915MHz band (only in the US) and 1 channel in the 868MHz band (EU and Japan). The 2450MHz band allows higher datarate and offers more channels than the other bands and thus is well suited for sensor networks with high network load. Signaling in the 2450MHz band is based on orthogonal quadrature phase shift keying (O-QPSK) and direct sequence spread spectrum (DSSS). The chip rate equals 2 Mchip/s. One 4-bit symbol is mapped into a 32-chip PN sequence, resulting in a symbol period $T_S$ of 16µs, and a throughput of 250 kbps corresponding to a byte period $T_B$ of 32µs. The CC2420 IC implements the 2450GHz PHY and supports MAC functionalities. The transmitter and receiver have respectively a direct up-conversion and low IF I/Q architecture. The transmit power can be programmed from –15 to 0 dBm in 8 steps.

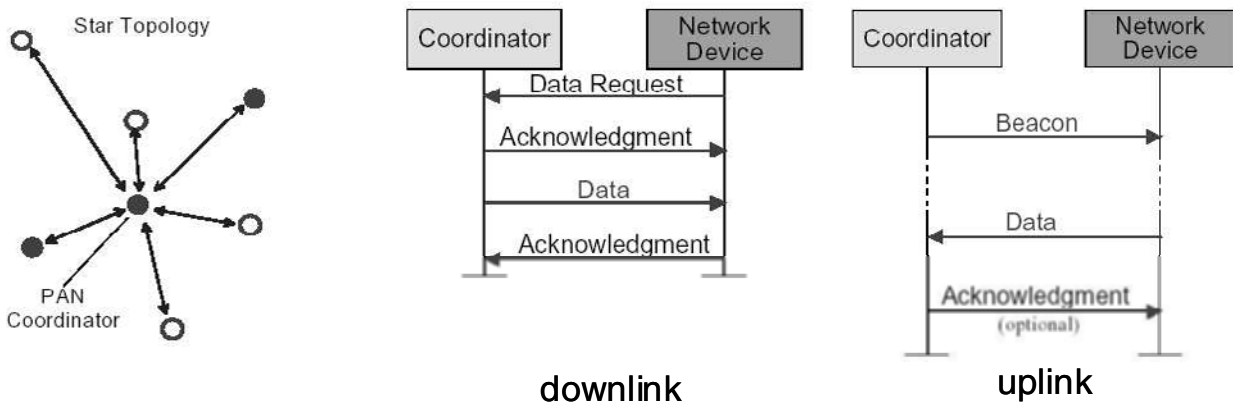

Figure 1: (a) Star topology in beacon mode. (b) Indirect transmission is used in the downlink while slotted CSMA/CA is used for the uplink

### Medium access control layer

Many possible network topologies can be built based on the 802.15.4 MAC. We focus only on 1-hop star networks (Figure 1a) where a *network coordinator* is elected. In a wireless microsensor network, the network coordinator can be the base-station. Communication from nodes to coordinator (uplink), from coordinator to node (downlink) or from node to node (ad hoc) is possible. In the following, we model uplink communication, which occurs more often than downlink or ad hoc communication in a network that gathers information from the environment and forwards it to the base-station.

In a star network, the beacon mode appears to allow for the greatest energy efficiency. Indeed, it allows the transceiver to be completely switched off up to 15/16 of the time when nothing is transmitted/received while still allowing the transceiver to be associated to the network and able to transmit or receive a packet at any time [10]. The beacon mode introduces a so-called superframe structure (Figure 2). The superframe starts with the beacon, which is a small synchronization packet sent by the network coordinator, carrying service information for the network maintenance and notifying nodes about pending data in the downlink. The inter-beacon period is partially or entirely occupied by the superframe, which is divided in 16 slots. A number of slots at the tail of the superframe may be used as *guaranteed time slot (GTS),* i.e. they are dedicated to specific nodes. This functionality targets very low latency applications but does not fit well in a dense sensor network since the number of dedicated slots would not be sufficient to accommodate several hundreds of node. In such conditions, it is better to use the contention access mode where the sparse data is statistically multiplexed.

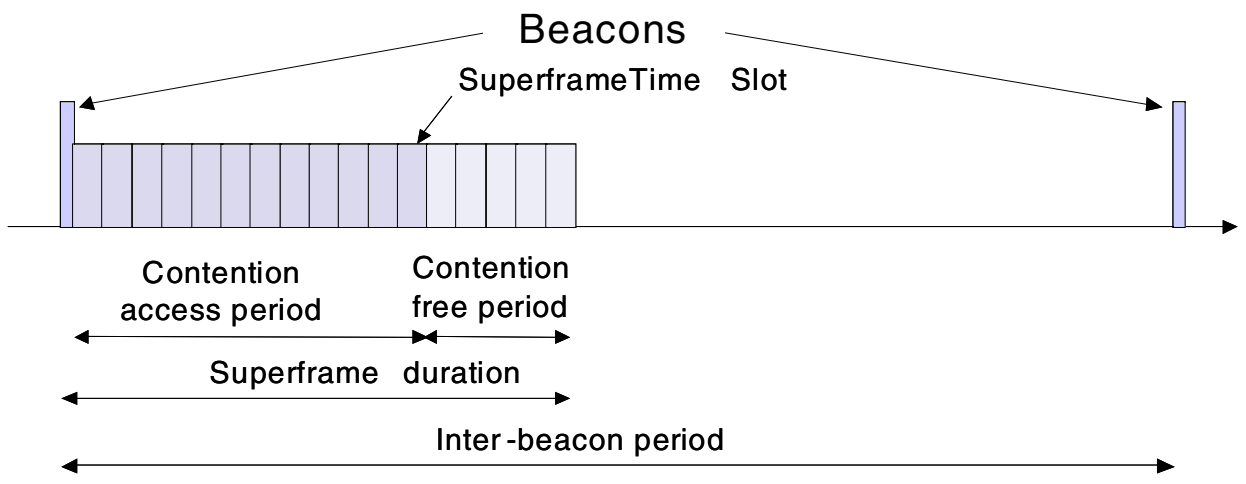

Figure 2: Superframe structure in beacon mode

In the contention access period, distributed channel accesses in the uplink are coordinated by a slotted carrier sense multiple access – collision avoidance (CSMA/CA) mechanism while indirect transmission is used in the downlink (Figure 1b). As we will see later, the CSMA/CA mechanism has a significant impact on the overall energy and performance of the uplink.

According to the slotted CSMA/CA algorithm [7], a node must sense the channel free at least twice before being able to transmit, this corresponds to the decrement of the so-called contention windows (CW). The first sense must be delayed by a random delay chosen between 0 and $2^{BE}-1$, where BE is the backoff exponent. This randomness serves to reduce the probability of collision when two nodes simultaneously sense the channel, assess it free and decide to transmit at the same time. When the channel is sensed busy, transmission may not occur and the next channel sense is scheduled after a new random delay computed with an incremented backoff exponent. If the latter has been incremented twice and the channel is not sensed to be free, a transmission failure is notified and the procedure is aborted. When a packet collides or is corrupted, it can be retransmitted after a new contention procedure.

The contention procedure starts immediately after the end of the beacon transmission. All channel senses or transmissions must be aligned with the CSMA slot boundaries that are separated by a fixed period of $T_{slot} = 20 \times T_s$.

As we will see later, the contention procedure introduces a significant overhead in energy consumption. Therefore a *Battery Life Extension* mode, where the backoff exponent is limited to 0-2 is supported by the 802.15.4 standard. However, in dense network conditions, this mode would results into an excessive collision rate. Hence, we are not using this feature in our experiment.

## 3. Characterization of the radio

To be able to assess the average power consumption of an 802.15.4 node in a network we must characterize the instantaneous power consumption of the transceiver when operating in and switching between states. The CC2420 transceiver supports four states:

1. *Shutdown*: The clock is switched off and the chip is completely deactivated waiting for a startup strobe
2. *Idle*: The clock is turned on and the chip can receive commands (for example, to turn on the radio circuitry)
3. *Transmit*
4. *Receive*

In the context of wireless microsensor networks, which are characterized by a very low transmission duty cycle, it has been shown that the transient energy when switching from one mode to another significantly impacts the total power consumption [11,12]. As we will see later, when considering the MAC, this effect becomes more significant. Hence, it is important to precisely characterize the transition time and energy between the transceiver states.

Steady state power, transient time and energy measurement have been made through the use of the Chipcon CC2420EM/EB evaluation board and the *SmartRF*$^{TM}$ *Studio* software [9]. Measurement results are summarized in Figure 3. The state transition energy has been evaluated by multiplying the transition time by the power in the arrival state. This is a worst-case assumption. Notice that the idle state power of 712µW is already 7 times higher that the average power goal of 100µW. To achieve lower power, the transceiver must enter the shutdown state when no action is required during a superframe. This is handicapped by the relatively long transition between shutdown and idle states (~1ms). To account for this delay, the chip must be preemptively turned on 1ms before the beacon. Fortunately, this transition requires a relatively low energy (691pJ). However, additional hardware is required to stay synchronized with the superframe, as the CC2420's clock is turned off in the shutdown state.

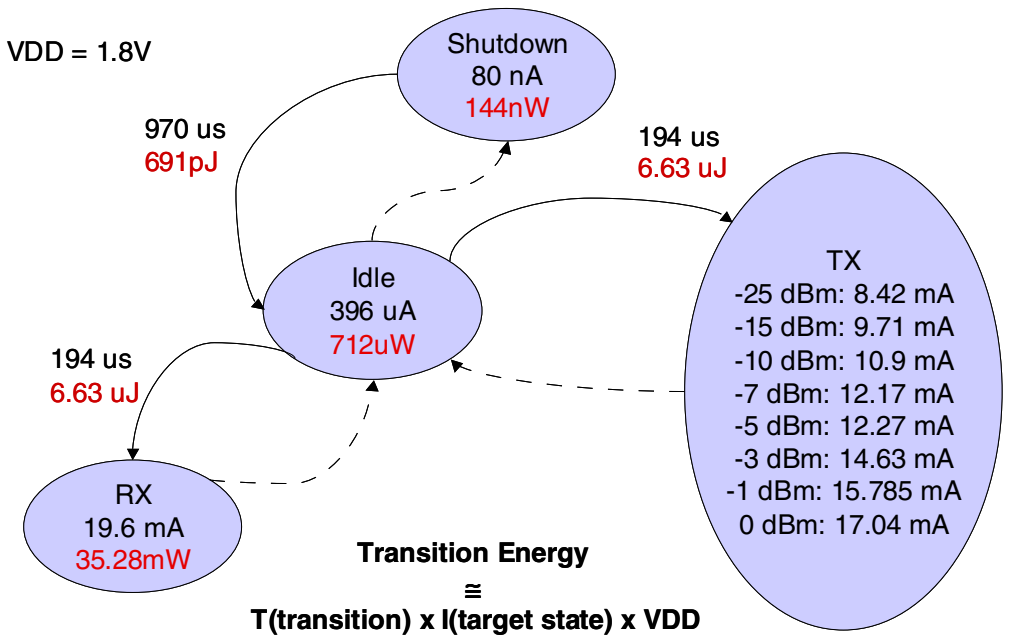

Figure 3: Steady state and transient power and energy measurement results

After characterizing the energy behavior of the transceiver, the next step is to examine the overall link performance. The bit error probability has been estimated on a testbench composed of a CC2420 transmitter wired to a second CC2420 in receive mode, through a set a calibrated attenuators. Using a wired channel allows one to precisely control the received power. The conditions of an additive white gaussian noise channel (AWGN) are reproduced to assess the packet error probability as a function of the received power. The assumption of an AWGN channel is valid as long as the channel is coherent during the transmission of a packet (slow fading). With the maximum packet size of 123 bytes transmitted at the gross rate of 250kbps, the packet transmission takes 4 ms, which is smaller than the coherence time encountered in the 2450GHz band without mobility issues [13]. The estimated bit error probability is plotted in Figure 4. An exponential regression is done leading to equation (1) where $Pr_{bit}$ is the bit error probability and $P_{Rx}$ the received power, equal to the transmit power $P_{Tx}$ minus the pathloss $A$.

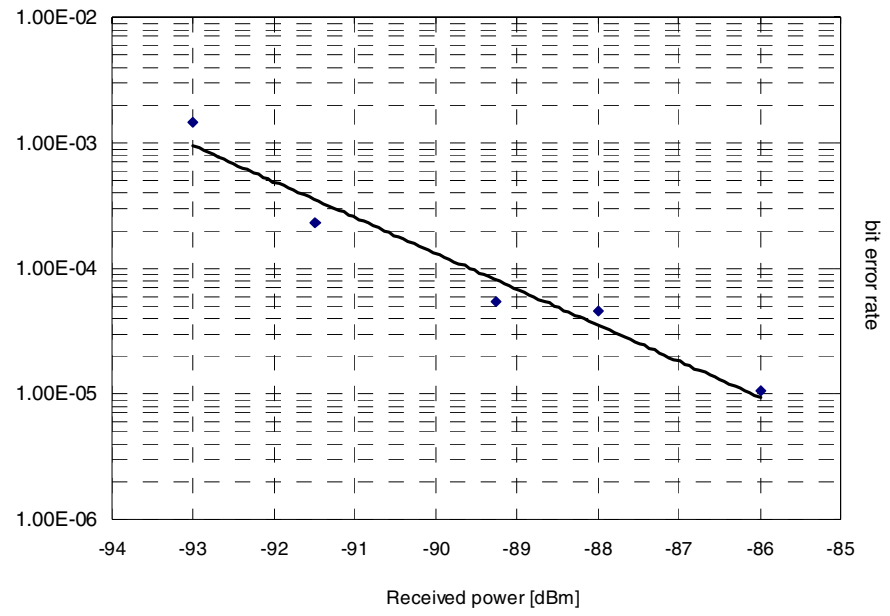

Figure 4: Bit error probability estimation results

$$Pr_{bit} = 2.35.10^{-30}.e^{-0.659.P_{Rx}} \quad (1)$$

$$P_{Rx} = P_{Tx} - A \quad (2)$$

## 4. Radio activation policy, link adaptation and average power consumption

***Transmission procedure and radio activation***

The data of Figures 3 and 4 are sufficient to characterize the performance and energy of the physical layer but they are not sufficient to compute the average power and the transmission reliability of a node in network conditions. Indeed, as sketched for the uplink in Figure 5, the medium access control procedures introduce a significant overhead. In the following, we assume that a node will attempt to transmit a single packet per superframe. To do so, it will first listen the beacon, after having preemptively turned on its radio in receive mode. After the beacon is received, the node can enter idle mode. As explained in section 2, the contention procedure requires at least two channel senses for clear channel assessment (CCA), which requires turning the receiver on. Between the CCAs, the receiver can return to the idle state. The node must stay in idle rather than shutdown because of the 1 ms delay to recover from the shutdown state. Once the channel is assessed clear twice, the transmission can start.

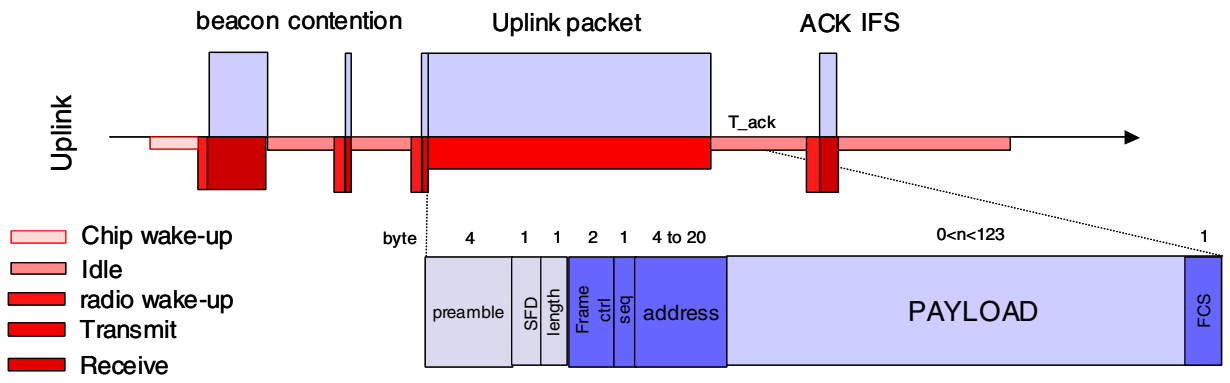

Figure 5: MAC overheads for the uplink

If the channel coherence time is sufficiently large (larger than a few packet transmissions), the transmit power can be selected as a function of the path loss measured during the reception of the beacon. Since the data rate is fixed, the best link adaptation policy at this level is channel inversion, i.e. keeping the receiver signal to noise ratio constant by compensating for the channel fading by increasing transmit power. In section 5, we compute the thresholds to switch from one power level to the other.

In addition to the modulated payload data, the transmitted packet consists of a preamble sequence to ease synchronization (corresponding to 4 bytes of data), a frame delimiter (1 byte) and 1 and 8 bytes of PHY and MAC service data, respectively. We assume that short (4 bytes) addresses are used. Let $L_o = 13$ be the total overhead in byte introduced by the PHY and the MAC. The time needed to transmit a packet is given by (3).

$$T_{packet} = (L_o + L) \times T_B \qquad (3)$$

As aforementioned, despite the CSMA/CA procedure, there exists a probability the packet collides with the transmission of another node. Also, the packet can be corrupted by bit errors due to noise. Therefore, a packet acknowledgment mechanism is implemented. If the packet is well received, a short acknowledgement packet is fed back to the transmitter after a minimum time $t^-_{ack} = 192\mu s$. The transmitter waits for such an acknowledgement for maximum $t^+_{ack} = 864\mu s$. If nothing is received, the transmitter repeats the transmission. The node can enter idle mode during the $t^-_{ack}$ period but must be in receiving mode between the end of $t^-_{ack}$. and the reception of the acknowledgement or until the end of $t^+_{ack}$. Since we assume that the node transmits only one packet per frame, it can shutdown after receiving the acknowledgement.

To compute both the average power consumption of a node and the probability a transaction fails, we still have to characterize the average duration of the contention procedure ($\overline{T}_{cont}$), the average number of CCAs ($\overline{N}_{CCA}$) done during this period, the residual probability of collision ($Pr_{col}$) and the probability a channel access failure is reported ($Pr_{cf}$). These quantities depend mainly of the network load ($\lambda$) – defined as the aggregate data rate relative to the maximum bandwidth – and the packet duration ($T_{packet}$). We have characterized those relations empirically by Monte-Carlo simulation of the contention procedure. Results for a network of 100 nodes per channel are depicted in Figure 6.

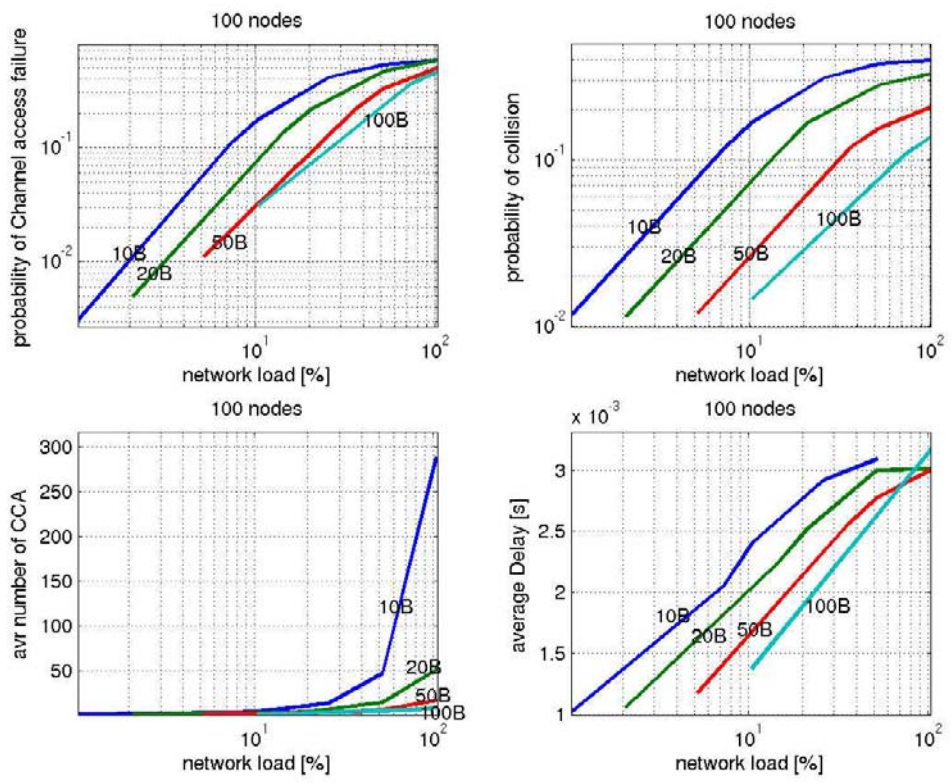

Figure 6: Behavior of the slotted CSMA/CA algorithm for different packet sizes (10 bytes, 20 bytes, 50 bytes and 100 bytes)

***Average power consumption***

To compute the average node power, we have to determine how long the node occupies each state. To account for the state transition energy, we add the transition delay ($T_{si}$ = 1ms: transition time between shutdown and idle; $T_{ia}$ = 194µs: transition time between idle and transmit/receive) to the corresponding active time. The expressions of the average time the node is in idle, transmit and receive modes when following the proposed activation policy are given in (4,5,6).

$$T_{idle} = T_{si} + \Pr_{cf} \times \overline{T}_{cont} + (1-\Pr_{cf}) \times \left[ \left( \sum_{i=1}^{N_{max}} P_{tr}(i) \times i + P_{tr}(>N_{max}) \times N_{max} \right) \times \left( \overline{T}_{cont} + t^{+}_{ack} \right) \right] \qquad (4)$$

$$T_{Tx} = (1-\Pr_{cf}) \times \left( \sum_{i=1}^{N_{max}} P_{tr}(i) \times i + P_{tr}(>N_{max}) \times N_{max} \right) \times T_{packet} \qquad (5)$$

$$T_{Rx} = T_{ia} + T_{beacon} + \Pr_{cf} \times \overline{N}_{CCA} \times T_{ia} + (1-\Pr_{cf}) \times \left[ \begin{array}{l} \left( \sum_{i=1}^{N_{max}} P_{tr}(i) \times i + P_{tr}(>N_{max}) \times N_{max} \right) \times \overline{N}_{CCA} \times T_{ia} \\ + \sum_{i=2}^{N_{max}} P_{tr}(i) \times (i-1) \times t^{+}_{ack} + (1 - P_{tr}(>N_{max})) \times T_{ack} \end{array} \right] \qquad (6)$$

$P_{tr}(i)$ is the probability $i$ transmissions are required to transmit a packet. The maximum number of transmissions, $N_{max}$, is limited to 5 in our investigation. $P_{tr}(i)$ can be computed by (7,8) where $Pr_{tf}$ is the probability of transmission failure (9) combining the probability of collision $Pr_{col}$ and the probability of transmission error $Pr_e$, which is computed as a function of the bit error probability ($Pr_{bit}$) and the total packet size ($L_{packet}$) minus the synchronization preamble (10).

$$P_{tr}(i) = \Pr_{tf}^{\,i-1} \times (1 - \Pr_{tf}) \qquad (7)$$

$$P_{tr}(>N_{max}) = 1 - \sum_{i=1}^{N_{max}} P_{tr}(i) \qquad (8)$$

$$\Pr_{tf} = 1 - (1 - \Pr_{col}) \times (1 - \Pr_{e}) \qquad (9)$$

$$\Pr_{e} = 1 - (1 - \Pr_{bit})^{(L_{packet}-4) \times 8} \qquad (10)$$

To compute the average power, $T_{idle}$, $T_{Rx}$ and $T_{Tx}$ must be multiplied by the steady state power in the corresponding mode ($P_{idle}$, $P_{Rx}$ and $P_{Tx}$) and divided by the inter-beacon period (11). The leakage power when the chip is shutdown is neglected. The inter-beacon period ($T_{ib}$) is computed as a function of the minimum superframe duration ($T_{ib}^{min}$ = 15.36 ms) and the so-called beacon order ($BO$), which can be chosen between 0 and 15 (12).

$$P_{avr} = \frac{P_{idle} \times T_{idle} + P_{Tx} \times T_{Tx} + P_{Rx} \times T_{Rx}}{T_{ib}} \quad (11)$$

$$T_{ib} = T_{ib}^{\min} \times 2^{BO} \quad (12)$$

***Probability of transmission failure***

In addition to the average power consumption, the total probability of transmission failure can be computed. A transmission failure can be due to a channel access failure, which occurs with a probability $Pr_{cf}$, or if the packet cannot be transmitted after $N_{max}$ trials (probability: $P_{tr}(>N_{max})$). The transmission failure probability can be computed as:

$$\Pr_{fail} = 1 - (1 - \Pr_{cf}) \times (1 - P_{tr}(> N_{\max})) \quad (13)$$

## 5. Case study

With the model developed in section 4, it is now possible to study the energy efficiency of the IEEE 802.15.4 standard in the context of dense microsensor networks. We consider a scenario where 1600 nodes are uniformly distributed in a circular area around a base-station. Since 16 channels are available, 100 nodes are sharing the same channel. We will assume that each node attempts to transmit 1 byte of data every 8 ms, resulting in an effective data rate of 1kbps per node and 100kbps per channel.

***Link adaptation***

We assume that all the nodes are within communication range of the base-station, i.e. the pass loss between one node and the base-station is such that the received power when 0 dBm are transmitted is above the receiver sensitivity. Since the different nodes experience different path losses, to achieve maximum energy efficiency, they have to adapt their transmit power. To determine the energy-optimal thresholds to switch between transmit power levels, the total energy per transmitted bit is computed for the full range of path loss. We assume that if a transmission fails in a given superframe, the application will retry transmission in the next superframe. The average transmission delay and average energy per bit are hence given by:

$$delay = T_{ib} \times \frac{1}{1 - \Pr_{fail}} \quad (13)$$

$$energy = \frac{P_{avr} \times delay}{L_{data} \times 8} \quad (14)$$

The results for a packet size of 120 bytes are depicted in Figure 7. Power level thresholds (circles) correspond to the crossing of energy-pathloss curves for the different transmit power levels. It can be seen that the thresholds are independent of the network load. The transmission is efficient for path losses up to 88 dB. The energy per bit ranges from 135nJ/bit for a pathloss lower than 55dB to 220nJ/bit for a pathloss of 88 dB. Hence, adaptation of the transmit power can save up to 40% of the total energy.

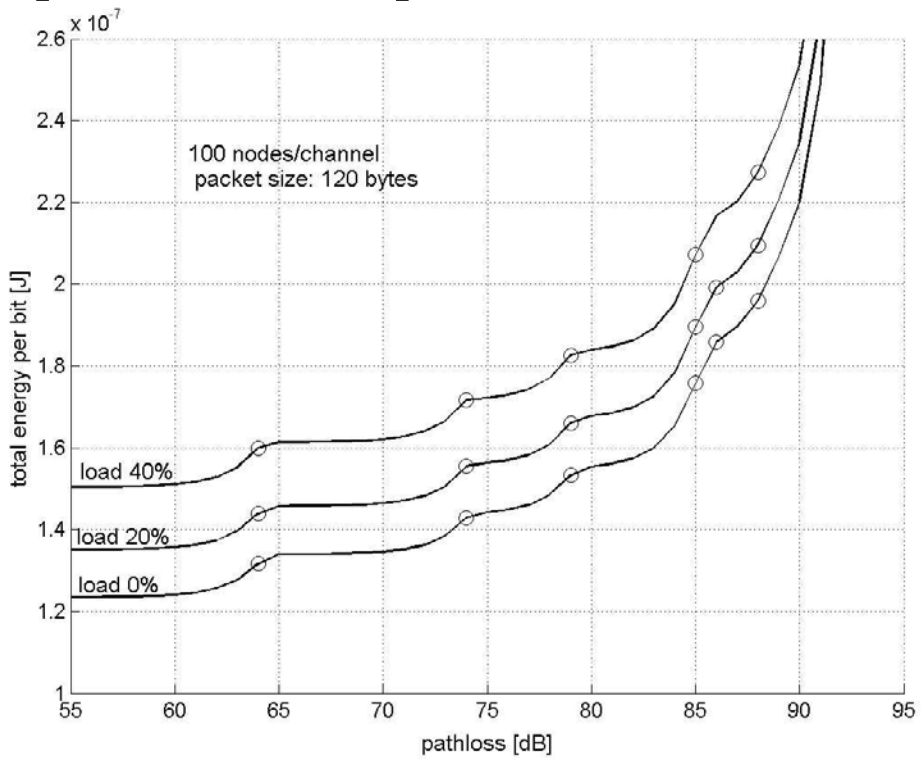

Figure 7: Optimal energy per bit for different path loss and network loads

***Packet size***

When the optimal transmit power is known, it is interesting to determine which packet size leads to the minimum energy per bit. On one hand, small packets require the same MAC overhead as large packets, which increases their energy per useful bit. However, large packets are more subject to transmission error, and hence require retransmission more often. In addition, when network load is high, large packets will increase the channel access failure probability. Intuitively a tradeoff is expected. However, as depicted in Figure 8, the energy per bit decreases monotonically up to a packet payload size of 123 bytes, which is the maximum possible in 802.15.4. Reaching the optimum requires a larger packet size.

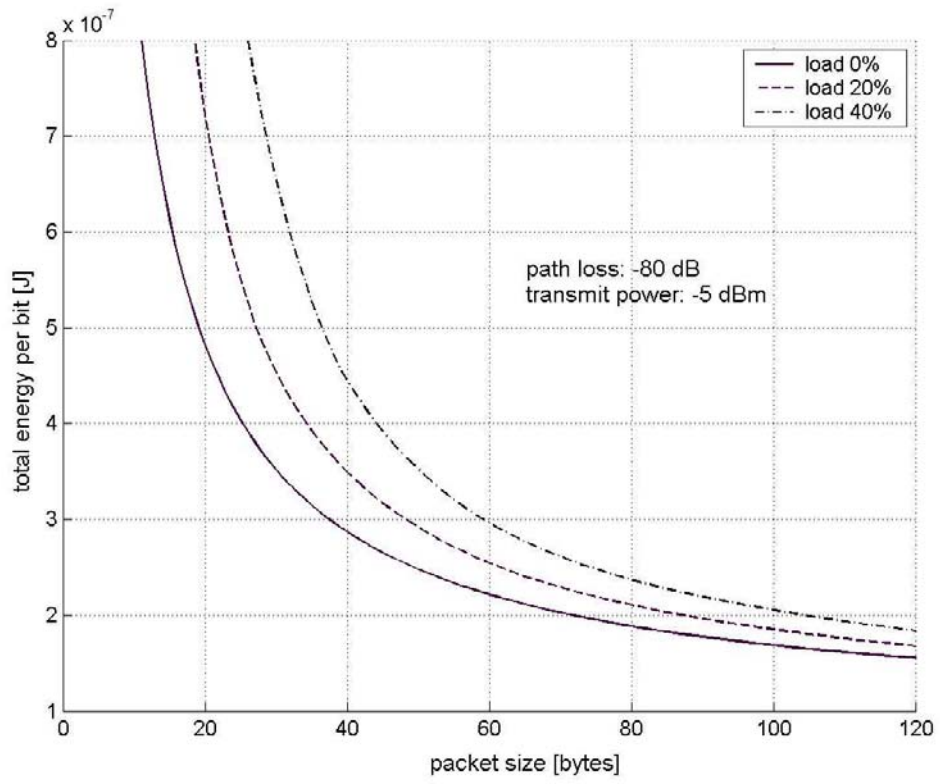

Figure 8: Impact of the MAC overhead at different network loads

We hence choose for this case study a packet size of 120 bytes. Gathered data is buffered until 120 bytes are accumulated. Thus, each node attempts to transmit a packet every 960ms. We set the beacon order to 6, so that one packet per node is transmitted during each superframe. This corresponds to a load of 42% in each channel. As-suming that the path loss is distributed uniformly between 55 and 95 dB, one can compute using the model presented in section 4 that the average power equals 211μW with a delivery delay of 1.45s and a probability of transmission failure of 16%.

***Power breakdown***

Interestingly, the calculated power consumption is close but still over the existing 100μW constraint of energy scavenging. To lower power consumption in future designs, it is valuable to know the energy breakdown of the node. Figure 9 presents the energy breakdown between the different phases of the protocol in our scenario. We notice that the effective transmission uses less than 50% of the total energy. 25% of the energy is spent during contention. This is due to the multiplicative effect of the CSMA/CA mechanism to the transceiver start-up energy. The acknowledgement mechanism uses 15% of the energy, mainly because of the necessity of activating the receiver during the acknowledgement waiting-time. 20% of the energy is spent for listening for the beacon.

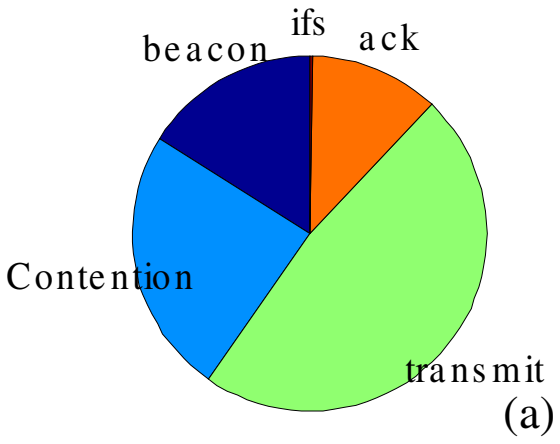

(a)

Shutdown: 98.77%
Idle: 0.47%
Transmit: 0.48%
Receive: 0.28%

(b)

Figure 9: Breakdown of the energy per bit (a) and time (b) spent in the different phases of the protocol

Based on the energy breakdown for the transceiver, one can see several key ways to improve the overall energy efficiency of sensor networks. Specific methods include reducing the transition time between states and designing a scalable receiver. Reducing the transition time between states by a factor two would decrease the total average power by 12%. Furthermore, a scalable receiver that offers a low power mode for sensing the channel and waiting for an acknowledgement frame has the potential of reducing the total average power by an additional 15%.

## 6. Conclusions and perspectives

The release of the IEEE 802.15.4 standard has been a major milestone in the transition of wireless microsensor networks from the research world to industrial applications. In this paper, we have studied how this standard can be used to support communication in dense, data-gathering networks. An energy-aware radio activation policy has been proposed and the corresponding average power consumption and transmission reliability have been analyzed as a function of important network parameters. The resulting model has been used to optimize the physical and medium access control layers parameters in a dense sensor network scenario. We have analyzed a sensor network of 1600 nodes transmitting 1 byte every 8 ms and calculated the average power consumption to be 211μW. To achieve this power figure, buffering is necessary in order to use the largest packet size allowed by the standard. Indeed, the energy per bit decreases monotonically with the packet size up to the maximum allowed size. Allowing larger packets would allow further energy efficiency improvement, at the cost of increased latency. It has been shown that in the considered scenario, less than 50% of the energy is used for actual data transmission. A significant percentage of energy is consumed during the contention procedure (25%) and waiting for an acknowledgement (15%). The overhead of the contention is mainly due to the receiver start-up energy when doing clear channel assessment. The acknowledgement overhead results from the receiver power consumption when waiting for an acknowledgment. Based on the energy breakdown, several ways to improve the overall energy efficiency are proposed. These physical level improvements combined with continued MAC optimizations will allow for energy efficient, self-powered sensor networks.

## 7. Acknowledgement

This work is the result of the collaboration between the Digital Integrated Circuits and Systems Lab at MIT and the Wireless Research group at IMEC, thanks to special credit from the Belgian National Science Foundation (FWO). Bruno Bougard is Research Assistant at the FWO. The author would like to thank Chipcon Inc. for donating the evaluation boards used in this experiment.

## References

[1] R.L. Ashok, D.P. Agrawal, “Next-generation wearable networks, “ *Computer, Vol. 36, Issue*: 11, Nov. 2003, pp.31 – 39.

[2] D. Estrin and R. Govindan, “Next century challenges: Scalable coordination in sensor networks,” in *Proc. Mobicom*, 1999, pp. 263-270.

[3] A.P. Chandrakasan et al, “Design considerations for distributed micro sensor systems,” in *Proc. CICC*, 1999, pp. 279-286.

[4] S. Roundy at al., “Energy Scavenging for Wireless Sensor Networks, with Special Focus on Vibration”, Springer, 2003

[5] R. Min et al., "Energy-centric enabling technologies for wireless sensor networks," IEEE Wireless Communications, Vol. 9, No. 4, pp. 28-39, Aug 2002

[6] J.M. Rabaey et al, “PicoRadio supports ad hoc ultra low power wireless networking,” *IEEE Computer*, vol.33, pp. 42-48, 2000

[7] http://standards.ieee.org/getieee802/download/802.15.4-2003.pdf

[8] G. Lu et al., “Performance Evaluation of the IEEE 802.15.4 MAC for Low-Rate Low-Power Wireless Networks”, in Proc. EWCN’04, April 2004

[9] http://www.chipcon.com/files/CC2420_Data_Sheet_1_2.pdf

[10] J. Zheng and M.J. Lee, “Will IEEE 802.15.4 Make Ubiquitous Networking a Reality?” IEEE *Communication Mag.*, Jun 2004

[11] Rex Min et al., “A Framework for Energy-Scalable Communication in High-Density Wireless Networks,” in *Proc. ACM ISLPED*, pp 36-41, Monterey, Aug. 2002

[12] E. Shih et al., “Design Consideration for Energy-Efficient Radios in Wireless Microsensor Networks,” in *Journal of VLSI Signal Processing*, vol. 37, pp 77-94, 2004

[13] S. Thoen et al., "Channel Time-Variance for Fixed Wireless Communications: Modeling and Impact," in *Proc. IASTED Wireless and Optical Communications Conference,* Banff, June 2002